# The three-step workflow: a pragmatic approach to allocating academic hospitals' affiliations for bibliometric purposes

Andrea Reyes Elizondo[1], Clara Calero-Medina[1], Martijn S. Visser[1]

[1] *Centre for Science and Technology Studies (CWTS), Leiden University, Leiden, The Netherlands*

ORCID:

Andrea Reyes Elizondo: 0000-0002-5676-2122

Clara Calero-Medina: 0000-0001-8736-6571

Martijn S. Visser: 0000-0001-5987-2389

Email of corresponding author:

a.e.reyes.elizondo@cwts.leidenuniv.nl

**Author contributions**

ARE, MSV, and CCM contributed to the conception of the study and writing of the manuscript; ARE and MSV designed the models and workflow; while MSV was responsible for the analysis of the output.

**Conflict of interest**

The authors have no conflicts of interest to declare that are relevant to the content of this article.

**Funding**

This work was supported by RISIS – Research Infrastructure for Research and Innovation Policy Studies, an EU FP7 Research Program Project [grant agreement no: 313082].

**Acknowledgements**

The authors wish to thank the Leiden Ranking team, especially the A-TEAM: Zeynep Anli, Maia Francisco Borruel, and Sonia Mena Jara.

## Abstract

This paper presents a method for classifying the varying degrees of interdependency between academic hospitals and universities in the context of the Leiden Ranking. A key question for ranking universities is whether or not to allocate the publication output of affiliated hospitals to universities.

Hospital nomenclatures vary worldwide to denote some form of collaboration with a university: academic hospitals, teaching hospitals, university hospitals, and academic medical centres do not correspond to universally standard definitions. Thus, rather than seeking a normative definition of academic hospitals, we are proposing a workflow that aligns the university-hospital relationship with one of three general models: full integration of the hospital and the medical faculty into a single organization; health science centres in which hospitals and medical faculty remain separate entities albeit within the same governance structure; and structures in which universities and hospitals are separate entities which collaborate with one another. This classification system provides a standard by which we can allocate publications which note affiliations with academic hospitals.

Our three-step workflow effectively translates the three above-mentioned models into two types of instrumental relationships for the assignation of publications: "associate" and "component". When a hospital and a medical faculty are fully integrated or when a hospital is part of a health science centre, the relationship is classified as component. When a hospital follows the model of collaboration and support, the relationship is classified as associate. The compilation of data following these standards allows for a more uniform comparison between worldwide educational and research systems.

## Keywords

Academic Hospitals; Universities; University Rankings; Bibliometrics; Organizational affiliations.

# 1. Introduction

Determining a universal definition for what constitutes a university is a challenging task (Birtwistle, 2003; Rochford, 2006) which is complicated by internationalization (Altbach, 2004), rankings (Dill, 2006), strategic mergers (Labi, 2011), and the evolving roles and social expectations placed on universities (Denman, 2005). Likewise, characterizing the relationships between universities and their affiliated organizations can be problematic given the different degrees of interdependency that exist between them. This is particularly evident in the case of academic hospitals (Barrett, 2008).

Not only do these institutions tend to follow particular national systems, but they also evolve constantly in response to economic factors and sociopolitical expectations (Wartman, 2008). Building on Praal et al.'s (2009) research into the challenges of including affiliated institutes in a university's publication count, this paper describes the various academic hospital systems we have encountered and proposes a mode of classifying the university-hospital relationship in the context of the Leiden Ranking.

Rather than seeking a normative definition of "academic hospital", we are proposing a method of allocating publications that only mention affiliations with such hospitals. For the purpose of the Leiden Ranking, we distinguish between three general models of hospital-university relationships in order to measure the research output of universities through publications. This exercise is necessary because the organizational boundaries of universities are not always clear. In addition, the information provided under author affiliations may not always be accurate, for example when researchers mention their working address at the affiliated hospital but omit their university affiliation. In this respect, our methodology aligns with Hardeman's (2013) claim that university output cannot be objectively identified because there is no universal and objective definition of what constitutes a university. This is exacerbated by the fact that affiliations serve two general purposes: providing contact information for the authors and attributing credit to the employer(s).

## 1.1 Organizations involved

A wide variety of definitions exist for what constitutes a university, a hospital, and in particular an academic hospital. Below we outline the concepts that will appear in this article. As per the Leiden Ranking framework, we consider universities as higher education institutions that award degrees of the third cycle (Leiden Ranking, 2018) in accordance with the *Framework for Qualifications of the European Higher Education Area* (Bologna Working Group on Qualifications Frameworks, 2005).

In general terms, a hospital is perceived as an institution for the provision of healthcare. In most countries, a hospital will be considered as academic when it is connected to a higher education institution in some form. Although a precise definition and clear organizational structure for academic hospitals does not currently exist, the attribute of "academic" indicates such hospitals have a tripartite mission of patient care, education, and research. For the purposes of our classification system, we are most concerned with the hospital's relationship to the university in the areas of education and research.

Before we move on, it is worth noting that academic medical centres can differ substantially in size and organizational complexity. In addition, these organizations are constantly changing. What was once a single academic hospital may now be an entire campus or an academic health science centre comprising several hospitals, clinics, laboratories, and even research centres; we do not differentiate between single academic hospitals and these larger structures. As aforementioned, what we seek is to employ the university-hospital relationship in the areas of education and research as an instrument to measure publication output. More specifically, we consider whether the publications mentioning only an affiliation with an academic hospital should be included or disregarded in the count of a university's output.

## 1.2 Literature review on academic hospitals

The majority of the literature on academic hospitals either examines the organizational aspect of such institutions or focuses on their various outcomes (Zinner & Campbell, 2009). Below we provide a non-exhaustive overview of the most common research topics.

Many studies discuss the various organizational structures of academic centres (Daniels & Carson, 2011; Davies et al., 2010; Barrett, 2008), including their governance (Gaynor et al., 2012; Griner & Blumenthal, 1998), and the effects of ongoing restructuring and mergers. Researchers also often highlight the complexity of processes that arises from specific policies (French et al., 2014) and budget constraints (Fuchs, 2013).

Other scientific literature focuses on the different expectations placed on academic hospitals due to their tripartite mission (Lozon & Fox, 2002) as well as on the changing roles and challenges that academic medical education faces (Gallin & Smits, 1998; Nonnemaker & Griner, 2001; Wartman et al., 2009). Related to this are studies analyzing the conflicting interests between universities, public health organizations, and industry (Brennan et al., 2006; Cohen & Siegel, 2005; Rothman & Chimonas, 2010; Washington et al., 2013). This is especially the case when higher education institutions must find befitting partnerships to fulfil the demands for innovation (Fish, 2013).

Several articles treat specifically the subject of academic science centres and health networks (Fish, 2013; Nicholson et al., 2015; Ovseiko et al., 2014; Wijgert, 2010). Such articles reflect upon the expanding and conflicting expectations that public institutions face in the wake of budget constraints and cost-reduction mergers, and the accountability issues that arise from these events.

Aside from some articles on resource utilization in relation to teaching activities (Sato & Fushimi, 2012), the area of education, a fundamental part of the tripartite mission of academic hospitals, has received little scholarly attention. When it does receive attention, rarely a distinction is made between the different types of education such as patient, professional, postgraduate, or undergraduate education. This lack of distinction in the literature is an indication of the organizational complexity whereby the different types of education provided at hospitals follow the priorities of different governmental departments, most often those of education and health. Therefore, the term "teaching" is generally used in the literature in its broadest sense.

Although most articles focus on regional or national cases, many authors devote attention to developing a general definition of what constitutes an academic hospital. However, the diversity of educational, research, and public health models, coupled with the ever-changing relationships between universities and hospitals, make organizational definitions challenging even at the national level. As several scholars note, if you have seen one [academic medical hospital or medical school], you have seen only one (Lozon & Fox, 2002).

## 2. Methodology

This section describes the three types of affiliation that we distinguish between an academic hospital and university. It then outlines how to identify the type of relationship by applying the three-step workflow and how to interpret the findings.

### 2.1 Relationship models of academic hospitals

Hospitals affiliated with universities are relevant elements of academic systems where education and research take place (Washington et al., 2013). Nevertheless, there are significant differences in the types of relationships between hospitals and universities both on an international and national level: on one end of the spectrum there are fully integrated hospitals and medical faculties, while on the other end there are non-integrated hospitals offering clinical placements for students. In between lies an array of structures at varying degrees of interdependency, all determined by national and regional systems.

Examining the organizational structures of university-affiliated hospitals, Wartman describes two prototypical models of academic health centres, while Levine considers five possible structures with varying degrees of interdependency (Wartman, 2015; Levine, 2002). For the purpose of the Leiden Ranking, we identify three general models for relationships between universities and hospitals:

1. full integration of the hospital and the medical faculty into a single organization;
2. health science centres in which hospitals and medical faculty are still separate entities, albeit within the same governance structure;
3. structures in which universities and hospitals are separate entities and collaborate with one another.

The case of fully integrated academic hospitals shows a strong degree of interdependency with a particular university, for example, when the medical faculty is located at the hospital or vice versa (see Figure 1). This model is found in most academic hospitals of Belgium, the Netherlands, Germany, and Switzerland.

**Figure 1: Full integration model**

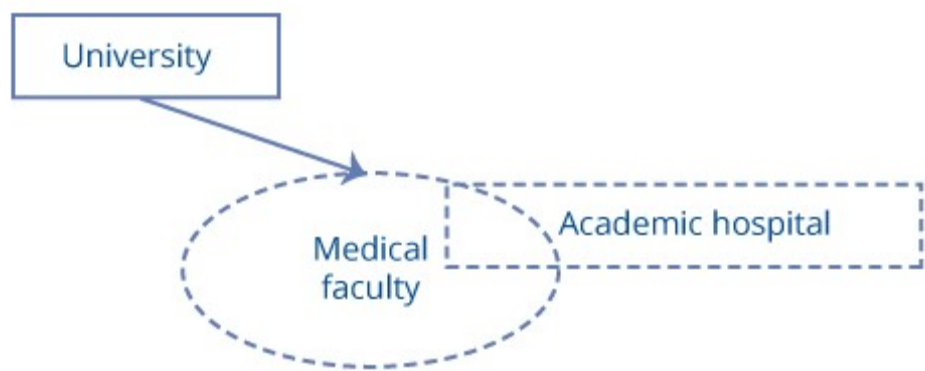

Source: authors

The system of Health Science Centers also presents a strong degree of interdependency between academic hospitals and universities but the structure differs from the full integration model. In the full-integration model, the medical faculty and hospital are often the same institution. Health science centres, on the other hand, are organizations that coordinate the medical faculty, the various patient-oriented medical facilities, and the research centres (see Figure 2). They can fall directly under the university or stand as a separate legal entity. Health science centres exist predominantly in the United States of America (e.g. the *University of New Mexico Health Sciences Center*).

**Figure 2: Health science centres**

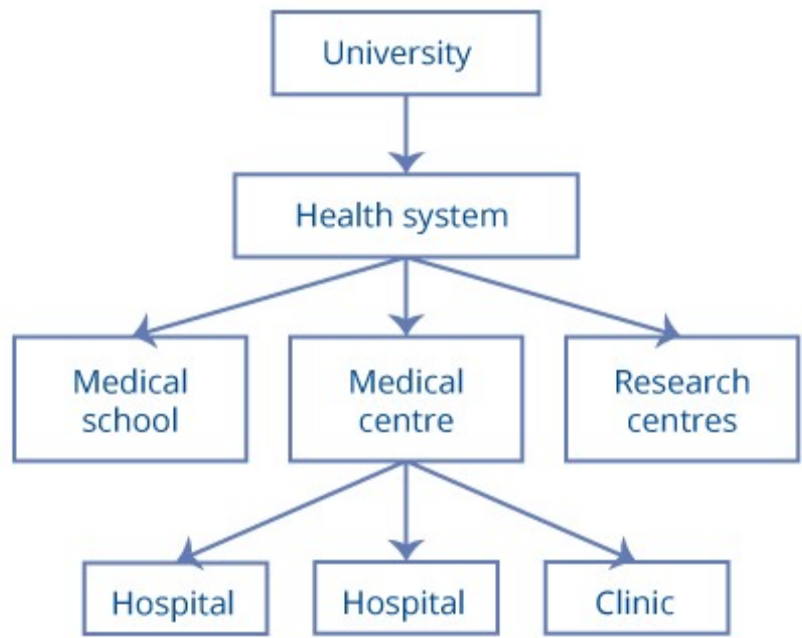

Source: authors

It is important to note the difference between health science centres (as modelled in Figure 2) and academic health science networks such as the *Oxford Academic Health Science Network* in the United Kingdom. These health networks refer to partnerships between one or more universities, local governments, and industries that seek to improve innovation delivery times and patient services (Fish, 2013). We have found that health networks are rarely mentioned in publications; furthermore, we have found that they pursue varied interests and do not focus primarily on undergraduate education. Other networks such as the *Copenhagen University Hospital* in Denmark and the *University Health Network* in Canada refer to collaboration agreements between the regional health authorities and a university without institutionalized integration between the medical faculty and all the hospitals in said regions.

The third model refers to a relationship of collaboration and support between hospital and university in which the collaboration is structural rather than incidental: there are some shared activities between the two, but no full integration of the medical faculty into the hospital (see Figure 3). These types of relationships are the most common worldwide.

**Figure 3: Collaboration and support**

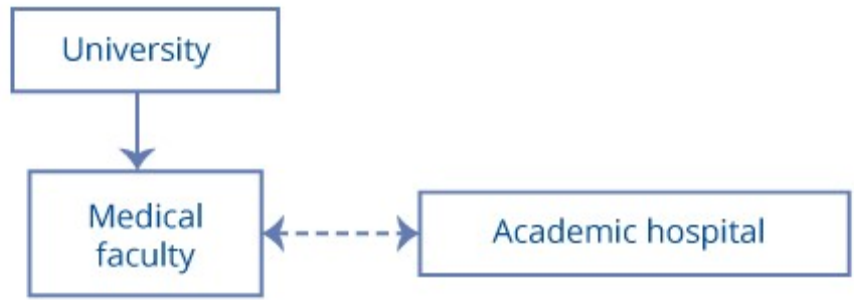

Source: authors

## 2.2 The three-step workflow: identifying the relationship

Around the world, hospitals are referred to as "academic hospital", "teaching hospital", "university hospital", "third-tier hospital" and "academic medical centre" (Shi et al., 2021; Sato & Fushimi, 2012) to denote some form of collaboration with a university; however, these nomenclatures do not constitute universally accepted definitions of specific types of hospital-university relationship, and thus can denote different things in different countries or even regions of the same country). Additionally, the level of collaboration between a university and a hospital may vary substantially between countries and regions, further rendering such labels unsuitable for global comparisons.

Sometimes the relationship between a hospital and a university may seem clear according to local perceptions. However, such local understandings of the interconnectedness between hospitals and universities cannot be used for international comparisons as they are not sufficiently consistent and demand specific local knowledge.

Following our previous work on academic hospitals (Praal et al., 2009), we at the Leiden Ranking have developed a three-step workflow to classify the relationship between universities and their affiliated hospitals. This workflow effectively transposes the three above-noted relationship models into the two types of instrumental relationships that the Leiden Ranking employs for the allocation of the publication output of academic hospitals.

We distinguish between academic hospitals that are associated with universities ("associate") and academic hospitals that can be considered as part of the university ("component").

When a hospital and a medical faculty are fully integrated or when a hospital is part of a health science centre (see Figures 1 and 2) the relationship is classified as component. When a hospital follows the model of collaboration and support (see Figure 3) the relationship is classified as associate. Although distinct in their organizational structures, the full-integration and health science centre models suggest a strong integrated relationship concerning the tasks of education and research. In both cases the medical faculty is difficult to separate from the hospital; therefore we treat them in the same way. In the collaboration and support model this is not the case.

Although the proposed scheme cannot do justice to the myriad forms of integration that might exist, it seeks to address the complexity of the relationship between academic hospitals and universities. The scheme is pragmatic and focuses on three aspects that can aid in evaluating the degree of interdependence: legal status, shared educational mandate, and level of physical integration (see Figure 4).

**Figure 4: Three-step workflow**

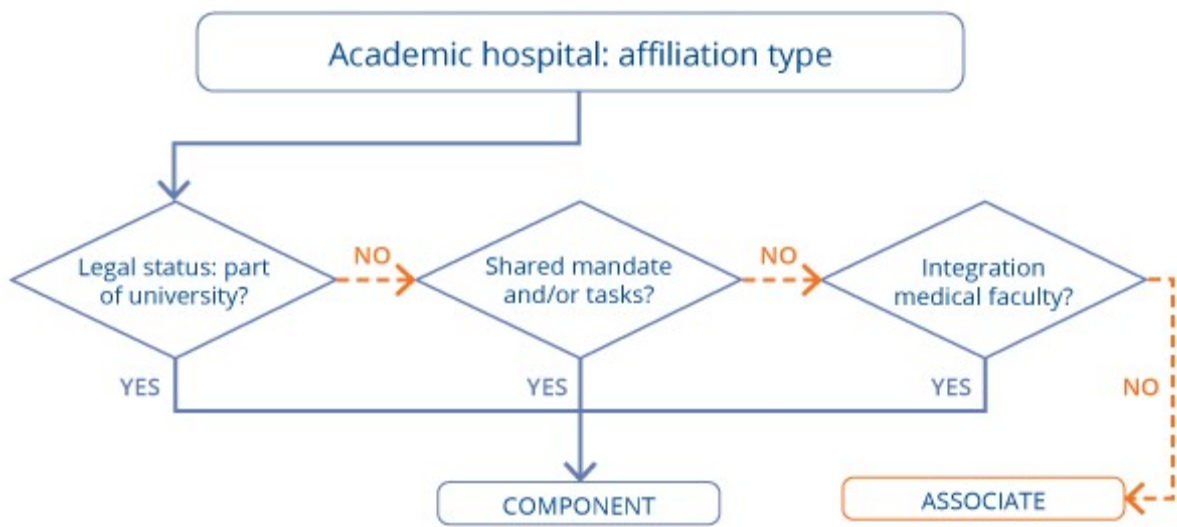

Source: authors

### 2.2.1 First step

First, a hospital's legal status is examined. If the ownership lies with the university or a foundation owned by or related to the university, the hospital is considered a component. In many cases, ownership can be easily determined for a whole country or region given specific health services regimes. For example, in the United Kingdom, hospitals and trusts that collaborate intensively with universities are owned by the *National Health Service*. In Poland, hospitals strongly connected to universities became part of the university structure by law in 2001. Thus according to this first step, hospitals in the United Kingdom cannot be considered as components, while academic hospitals in Poland should be considered as components of their universities.

For the first step in the workflow, not only the direct legal ownership is considered but also whether the hospital is part of a university health system. We find many cases like this in the United States of America. For example, the *Duke University Health System* comprises three hospitals, a medical faculty, and a series of outpatients' clinics. *Northwestern Medicine* has a similar structure that comprises various hospitals, a medical school, and a comprehensive cancer centre. Notably, an organizational structure like that of Duke or Northwestern is significantly different from a health science network as mentioned above, where the affiliated university exercises no direct control over the management of hospitals, clinics, and medical faculties.

Some cases present a seemingly opposite ownership situation that nevertheless results in a classification as component. For example, the *Mount Sinai Health System* in the United States of America owns the *Icahn School of Medicine*. This medical school was previously considered as part of *City University of New York* (1968–1998) and *New York University* (1999–2007) because these universities were granting its degrees until 2010; since then, *Mount Sinai* has been accredited as an independent degree-granting institution. Technically, the medical school belongs to the hospital group, yet for the Leiden Ranking's practical purposes (publication count) the hospital group is seen as a component of the medical school.

### 2.2.2 Second step

The second step in the workflow examines a hospital's mandate. Given the Leiden Ranking's focus on universities, the tasks considered here are core-curriculum education and research. In general, every hospital affiliated with a university will conduct research, except for smaller clinics which solely provide clinical placements. A more difficult function to determine is the type of education provided by a hospital. Over all, we have found that hospitals provide different levels of education ranging from patient education (for example on nutrition), specialist training or continuous education, to core-curriculum education for undergraduate students. On some occasions, a hospital has a clear mandate to aid a university in the provision of undergraduate education, in which case the hospital is considered to act as a component of the university.

The mandate of a hospital is not always publicly available, although for some countries it is part of the national or regional legislation. In a few cases, the mandate is part of a hospital's mission statement. For example, the *Centre hospitalier universitaire de Poitiers* in France mentions two types of education under its mission. The first type is referred to as "prevention and education", and targets patients as well as professionals. The second type concerns education for students of the University of Poitiers' medical faculty, including undergraduate education. In this case, although there is no mention of an explicit mandate, the hospital's mission does seem to indicate that it should be considered as a component. In instances like this wherein the mandate is not sufficiently clear, a full assessment following the three-step workflow is necessary in order to definitively determine whether the university-hospital relationship is associate or component.

In some countries, we have observed a clear pattern in which public academic hospitals provide core education. However, it is possible that the wave of public-services privatization in those countries may influence the extent of educational tasks provided by hospitals, as privatized hospitals tend to become less involved with education. In Germany, for example, the majority of large academic hospitals are traditionally public and involved with core education, but as some of these hospitals have been recently privatized, not all German hospitals can be automatically considered as components of universities.

Often, information about hospitals' mandates is not available at all. For example, the *Turku University Hospital* is owned by the *Hospital District of Southwest Finland* and seems to only cover continuous education for specialists. For such cases, which are the majority, the assessment requires progressing to the next step of the workflow.

### 2.2.3 Third step

The third and final step of the workflow examines to what extent a hospital and medical faculty are integrated. This step covers two major elements: the physical location of the faculty and the publication behaviour of authors mentioning the hospital affiliation. For the former we consider whether the hospital and the medical faculty have shared locations. For the latter we examine whether the authors publishing for the hospital are members of the medical faculty.

When a hospital shares locations and staff with a medical faculty, it is considered a component of the university. As in the previous steps, the outcome of this review is rarely a binary yes-no. In some cases, most departments of the medical school will be located on the hospital grounds, indicating a tight physical integration. Yet what we often find is that only one or two departments out of many are located at the hospital: usually pediatrics, gynaecology, clinical medicine, or nursing. In those cases, we do not consider this as a tight physical integration.

With this complexity in mind, the third step also considers the publication behaviour of authors using the hospital affiliation: we examine a sample of publications comprising authors using the hospital affiliation without mentioning the university to which the hospital is supposedly affiliated. We then investigate whether those authors are formally employed by the university. It is a sign of strong interdependency when the majority of authors publishing for a hospital are also faculty members at its affiliated university.

*2.2.4 Evaluating the three steps*

When the workflow shows a low level of interdependency, we classify the relationship of the hospitals as that of associate to the university. As an example, a hospital owned by the regional health authority, without a clear mandate for core-curriculum education, with only one university department located on the hospital premises, and few authors who hold a faculty position, would be considered as associate.

It is worth emphasizing that this is not a straightforward process: often it is difficult or impossible to retrieve all the information necessary to complete the workflow, and even with all of the information, the results of the assessment may be ambiguous. Thus the decision of classifying the relationship as associate or component typically depends on a combination of the available information. Lastly, because of its resource-intensive nature, the assessment is only carried out for academic hospitals whose publication output surpasses a specific threshold.

# 3. Evaluating the relationship - difference in numbers

Determining the relationship model can have a substantial impact on the publication counts for universities (as shown in Table 1). For example, the *University Hospital Zurich* (UHZ) in Switzerland is affiliated with the *University of Zurich* (UZ). The hospital is a legal entity on its own and although its mandate also includes educational responsibilities, it is not clear if those responsibilities extend to core-curriculum education. Nevertheless, the medical faculty and the hospital present a tight integration regarding physical locations and author affiliations:

many departments from the medical faculty are housed in hospital buildings, and the majority of authors who publish for the hospital hold a position at the university. University employees should include the university's address on their publication, yet despite efforts from university management encouraging authors to mention their affiliation, some do not. This is significant because if UHZ is treated as a component of UZ, its publication count increases by 23% (based on articles and reviews between 2014–2017), which highlights the importance of our categorizing exercise.

These differences in counting publications output are also evident in other cases. For example in Italy, the *Azienda Ospedaliero-Universitaria di Parma* raises the *University of Parma*'s publication count by 16%. In some cases, the contrast can be even more significant: in France, classifying the *Centre Hospitalier Régional Universitaire de Tours* as a component of the *Université de Tours* raises the publication count by 31%. Not all instances show such a large contrast, although the difference can still be notable. The *Ottawa Hospital* in Canada (which includes the *Children's Hospital of Eastern Ontario*) raises the publication count of the *University of Ottawa* by 13%. In Greece, the *University Hospital of Ioannina* raises the publication count of the *University of Ioannina* by only 8%, while the *University Hospital Olomouc* in the Czech Republic raises the count of *Palacký University Olomouc* by 7%.

Notably, the discrepancy on publication assignations is not present for all universities. For example, *Rovira i Virgili University* in Spain sees its count rise by 1% through the publications of the *Hospital Universitario Joan XXIII de Tarragona*, while the *University Hospital St. George* raises the count of the *Medical University of Plovdiv* in Bulgaria by 4%. In some cases, the differences are completely negligible, such as with the *University of Buenos Aires* in Argentina. The *Hospital de Clínicas "José de San Martín"* shows the characteristics of a component to the university, yet its publication output is so miniscule that it only raises the university's count by 0.3%.

| Organization | Separate Output | Combined Output | Percentage |
|---|---|---|---|
| Université de Tours | 1071,95 | 1561,36 | 31,34 |
| Centre Hospitalier Régional Universitaire de Tours | 489,41 | | |
| University of Zurich | 6449,56 | 8370,91 | 22,95 |
| University of Zurich Hospital | 1921,35 | | |
| University of Parma | 2066,87 | 2461,79 | 16,04 |
| Azienda Ospedaliero-Universitaria di Parma | 394,92 | | |
| University of Ottawa | 5733,92 | 6570,10 | 12,73 |
| The Ottawa Hospital | 557,41 | | - |
| Children's Hospital of Eastern Ontario | 278,77 | | - |
| University of Ioannina | 1183,65 | 1293,01 | 8,46 |
| University Hospital of Ioannina | 109,36 | - | - |
| Palacký University Olomouc | 1824,34 | 1969,87 | 7,39 |
| University Hospital Olomouc | 145,54 | - | - |
| Medical University of Plovdiv | 134,51 | 140,24 | 4,09 |
| University Hospital St. George | 5,73 | - | - |
| Rovira i Virgili University | 1695,18 | 1720,21 | 1,46 |
| Hospital Universitario Joan XXIII de Tarragona | 25,04 | - | - |
| University of Buenos Aires | 4089,15 | 4102,19 | 0,32 |
| Hospital de Clínicas "José de San Martín" | 13,04 | - | - |

Table 1. Overview of the separate output (articles and reviews 2014–2017) per university and hospital as well as their combined output. The percentage shows the increase of publication count of the university when the hospital is included as a component.

## 4. Advantages and limitations of the approach

As mentioned above, this paper does not seek to provide a normative definition of an academic hospital rather than a pragmatic process for assigning publications with only a hospital address in university rankings. The three-step model allows for a standard approach that can facilitate comparison on the level of integration between hospitals and universities on a global scale in a fair and transparent manner.

Further, by managing the affiliation relationships through linkages instead of automatically cleaning hospital publications as those of the university, data can be disaggregated and analyzed at different levels and combinations. Likewise, the university-hospital linkage is not an obstacle to recognize other types of collaboration between those same institutions. For example, an academic hospital may be an associate of a university yet collaborate tightly with it on a research lab. If the research lab has a significant publication output it can be created as unit of analysis which is component to both the university and the academic hospital. Thus the different collaboration paths can be recognized without focusing solely on the output of one organization (Hicks & Katz, 1996).

The approach is not without issues, especially as the information necessary to assign the type of affiliation is not always available. Owing to the different information-seeking paths and a non-binary decision process, the model cannot be applied automatically. Being labour and time intensive may limit its application by other research organizations. Nevertheless this model is a step in an ongoing research on publication patterns of hospitals and universities.

As the model is not normative, we do not foresee any sort of policy implications for academic hospitals and universities. The risk that some organizations or regions may game the rankings by specifically rewriting mandates is a possibility. However, the implications for the organizations themselves would be considerable as such change would involve a whole administrative and budgetary adjustment. At the same time, being transparent on the processes to categorize organizations and their affiliations may help universities and academic hospitals to communicate their relationship more clearly to the outside world.

## 5. Conclusion

Building upon Praal et al.'s (2009) discussions on how to delimit universities with regard to academic hospitals for publication count, the Leiden Ranking proposes a characterization and methodology that allows us to evaluate the degree of integration between these two types of organizations. Although the methodology presented herein is somewhat arbitrary, employing these standards to compile data allows for a more balanced and fair comparison between worldwide educational and research systems, without losing disaggregated data at different levels.

The Leiden Ranking's current approach has some limitations. Firstly, it requires a resource-intensive assessment process. We have sought to streamline this process by setting a threshold of publications before submitting a hospital to the three-step review. Secondly, examining the relationships between universities and hospitals in countries with non-Latin-based alphabets poses a challenge and can be a disadvantage for those organizations that do not have comprehensive faculty websites in English. Nevertheless, the model provides a standard approach for evaluating integration between these organizations.

To conclude, we would like to add that organizational structures will continue evolving in order to better respond to public and government expectations of accountability and efficiency, requiring our methodologies to evolve as well. The experience and knowledge acquired through the in-depth evaluation of academic hospitals following the three-step model, will allow us to follow these changes better.

**REFERENCES**

Altbach, P. G. (2004). Globalisation and the University: Myths and Realities in an Unequal World. *Tertiary Education and Management, 10*(1), 3–25. https://doi.org/10.1023/B:TEAM.0000012239.55136.4b

Barrett, D. J. (2008). The Evolving Organizational Structure of Academic Health Centers: The Case of the University of Florida. *Academic Medicine: Journal of the Association of American Medical* Colleges, 83(9), 804–808. https://doi.org/10.1097/ACM.0b013e318181d054

Birtwistle, T. (2003). What Is a "University"? (The English Patient)'. *Education and the Law, 15*(4), 227–236. https://doi.org/10.1080/0953996042000182138

Bologna Working Group on Qualifications Frameworks (2005). *A Framework for Qualifications of the European Higher Education Area*, (pp. 59–74). Copenhagen: Ministry of Science, Technology and Innovation.

Brennan, T. A., Rothman, D. J., Blank, L., Blumenthal, D., Chimonas, S. C., Cohen, J. J., Goldman, J., Kassirer, J. P., Kimball, H., Naughton, J., & Smelser, N. (2006). Health Industry Practices That Create Conflicts of Interest: A Policy Proposal for Academic Medical Centers. *JAMA, 295*(4), 429–433. https://doi.org/10.1001/jama.295.4.429

Cohen, J. J., & Siegel, E. K. (2005). Academic Medical Centers and Medical Research: The Challenges Ahead. *JAMA, 294*(11), 1367–1372. https://doi.org/10.1001/jama.294.11.1367

Daniels, R. J., & Carson, L. D. (2011). Academic Medical Centers–Organizational Integration and Discipline through Contractual and Firm Models. *JAMA, 306*(17), 1912–1913. https://doi.org/10.1001/jama.2011.1606

Davies, S. M., Tawfik-Shukor, A., & Jonge de B. (2010). Structure, Governance, and Organizational Dynamics of University Medical Centers in the Netherlands. *Academic Medicine, 85*(6), 1091–1097. https://doi.org/10.1097/ACM.0b013e3181dbf915

Denman, B. D. (2005). What Is a University in the 21st Century?. *Higher Education Management and Policy, 17*(2), 9–28. https://doi.org/10.1787/hemp-v17-art8-en

Dill, D. D. (2006, September 9). *Convergence and Diversity: The Role and Influence of University Rankings* [Keynote presented at the Consortium of Higher Education Researchers (CHER)] 19th Annual Research Conference, University of Kassel, Germany.

Fish, D. R. (2013). Academic Health Sciences Networks in England. *The Lancet, 381*(9882), e18–e19. https://doi.org/10.1016/S0140-6736(12)60866-6

French, C. E., Ferlie, E., & Fulop, N. J. (2014). The International Spread of Academic Health Science Centres: A Scoping Review and the Case of Policy Transfer to England. *Health Policy, 117*(3), 382–391. https://doi.org/10.1016/j.healthpol.2014.07.005

Fuchs, V. R. (2013). Current Challenges to Academic Health Centers. *JAMA, 310*(10), 1021–1022. https://doi.org/10.1001/jama.2013.227197

Gallin, J. I., & Smits, H. L. (1998). Managing the Interface Between Medical Schools, Hospitals, and Clinical Research. *Survey of Anesthesiology, 42*(1), 651–654. https://doi.org/10.1001/jama.1997.03540320053035

Gaynor, M., Laudicella, M., & Propper, C. (2012). Can Governments Do It Better? Merger Mania and Hospital Outcomes in the English NHS. *Journal of Health Economics, 31*(3), 528–543. https://doi.org/10.1016/j.jhealeco.2012.03.006

Griner, P. F., & Blumenthal, D. (1998). Reforming the Structure and Management of Academic Medical Centers: Case Studies of Ten Institutions. *Academic Medicine, 73*(7), 818–825. https://doi.org/10.1097/00001888-199807000-00025

Hardeman, S. (2013). Organization Level Research in Scientometrics: A Plea for an Explicit Pragmatic Approach. *Scientometrics, 94*, 1175–1194. https://doi.org/10.1007/s11192-012-0806-6

Hicks, D., & Katz S. (1996). Science policy for a highly collaborative science system. *Science and Public Policy* 23(1), 39–44. https://doi.org/10.1093/spp/23.1.39

Shi, H., Fan, M., Zhang, H., Ma, S., Wang, W., Yan, Z., Chen, Y., Fan, H., & Bi, R. Perceived health-care quality in China: a comparison of second- and third-tier hospitals. *International Journal for Quality in Health Care*, 33(1), 2021, https://doi.org/10.1093/intqhc/mzab027

Labi, A. (2011, January 2). University Mergers Sweep Across Europe. *The Global Chronicle*. Retrieved 26 March, 2019, from http://www.chronicle.com/article/University-Mergers-Sweep/125781

Leiden Ranking (2018) *Information, Universities*. The CWTS Leiden Ranking. Retrieved 18 February, 2019, from http://leidenranking.com/information/universities

Levine, J. K. (2002). Considering Alternative Organizational Structures for Academic Medical Centers. *AAMC Academic Clinical Practice, 14*(2) 2–5.

Lozon, J. C., & Fox, R. M. (2002). Academic Health Sciences Centres Laid Bare. *HealthcarePapers*, 2(3), 10–36. https://doi.org/10.12927/hcpap.2002.17202

Nicholson, K., Randhawa, J., & Steele, M. (2015). Establishing the SouthWestern Academic Health Network (SWAHN): A Survey Exploring the Needs of Academic and Community Networks in SouthWestern Ontario. *Journal of Community Health, 40*(5), 927–939. https://doi.org/10.1007/s10900-015-0015-3

Nonnemaker, L., & Griner, P. F. (2001). The Effects of a Changing Environment on Relationships between Medical Schools and Their Parent Universities. *Academic Medicine, 76*(1), 9–18. https://doi.org/10.1097/00001888-200101000-00007

Praal, F. E. W., Kosten, M. J. F., Calero-Medina, C., & Visser, M. S. (2013). Ranking Universities: The challenge of Affiliated Institutes. In: S. Hinze & A. Lottmann (Eds.) *Proceedings of the 18th International Conference on Science and Technology Indicators: Translational Twists and Turns: Science as a Socio-Economic Endeavour*, Berlin, IFQ, 2013, pp. 284–289.

Ovseiko, P. V., Heitmueller, A., Allen, P., Davies, S. M., Wells, G., Ford, G. A., Darzi, A., & Buchan, A. M. (2014). Improving Accountability through Alignment: The Role of Academic Health Science Centres and Networks in England. *BMC Health Services Research, 14*, 24. https://doi.org/10.1186/1472-6963-14-24

Rochford, F. (2006). Is There Any Clear Idea of a University?. *Journal of Higher Education Policy and Management, 28*(2), 147–158. https://doi.org/10.1080/13600800600750988

Rothman D. J., & Chimonas, S. (2010). Academic Medical Centers' Conflict of Interest Policies. *JAMA, 304*(20), 2294–2295. https://doi.org/10.1001/jama.2010.1714

Sato, D., & Fushimi, K. (2012). Impact of Teaching Intensity and Academic Status on Medical Resource Utilization by Teaching Hospitals in Japan. *Health Policy, 108*(1), 86–92. https://doi.org/10.1016/j.healthpol.2012.08.021

Wartman , S. A. (2015). The Academic Health Center in a Disrupted World. *The Pharos of Alpha Omega Alpha-Honor Medical Society, 78*(2), 2–9.

Wartman, S. A. (2008). Toward a Virtuous Cycle: The Changing Face of Academic Health Centers. *Academic Medicine*, 83(9), 797–799. https://doi.org/10.1097/ACM.0b013e318181cf8c

Wartman, S. A, Hillhouse, E. W., Gunning-Schepers, L., & Wong, J. E. L. (2009). An International Association of Academic Health Centres. *The Lancet, 374*(9699), 1402–1403. https://doi.org/10.1016/S0140-6736(09)61594-4

Washington, A., Coye, M. J., & Feinberg, D. T. (2013). Academic Health Centers and the Evolution of the Health Care System. *JAMA, 310*(18), 1929–1930. https://doi.org/10.1001/jama.2013.282012

Wijgert, J. van de. (2010). Academic Health Science Systems. *The Lancet, 375*(9728), 1782. https://doi.org/10.1016/S0140-6736(10)60807-0

Zinner, D. E., & Campbell, E. G. (2009). Life-Science Research within US Academic Medical Centers. *JAMA*, 302(9), 969–976. https://doi.org/10.1001/jama.2009.1265